\pdfoutput=1
\documentclass{article}

\usepackage{PRIMEarxiv}

\usepackage[utf8]{inputenc} 
\usepackage[T1]{fontenc}    
\usepackage{hyperref}       
\usepackage{url}            
\usepackage{booktabs}       
\usepackage{amsfonts}       
\usepackage{nicefrac}       
\usepackage{microtype}      
\usepackage{lipsum}
\usepackage{fancyhdr}       
\usepackage{graphicx}       
\graphicspath{{media/}}     
 
\usepackage{graphicx}
\usepackage{tikz-cd}
\usepackage[utf8]{inputenc}  
\usepackage{amssymb, amsmath}
\usepackage{amsthm}
\usepackage{enumitem}
\usepackage{hyperref}
\hypersetup{
    colorlinks=true,
    linkcolor=blue,
    filecolor=magenta,      
    urlcolor=cyan,
    citecolor=magenta,
    pdfpagemode=FullScreen,
    }

\theoremstyle{remark}

\theoremstyle{definition}
\newtheorem{theorem}{Theorem}[section]

\newtheorem{definition}{Definition}[section]

\newcommand{\cinf}{$\mathcal{C}^{\infty}$}
\newcommand{\contract}{\,\lrcorner\,}

\title{Integration of differential equations by \texorpdfstring{\cinf}--structures
\thanks{\textit{\underline{Citation}}: 
\textbf{Authors. Title. Pages.... DOI:000000/11111.}} 
}

\author{
  *A. J. Pan-Collantes, C. Muriel, A. Ruiz\\
  Departamento de Matem\'aticas\\ 
  Universidad de C\'{a}diz - UCA \\
  Puerto Real\\
  \texttt{\{antonio.pan@uca.es, concepcion.muriel@uca.es, adrian.ruiz@uca.es\}} \\
}

\begin{document}
\maketitle
\begin{abstract}
Several integrability problems of differential equations are addressed by using  the concept of \texorpdfstring{\cinf}--structure, a recent generalization of the notion of solvable structure. Specifically, the integration procedure associated with  \texorpdfstring{\cinf}--structures is used to integrate to a Lotka-Volterra model and several differential equations that lack sufficient Lie point symmetries and cannot be solved using conventional methods.
\end{abstract}

\section{Introduction}

Solvable structures appeared in the last decade of the 20th century as a generalization of the concept of solvable symmetry algebra \cite{olver86,ovsiannikovlibro,stephani,blumanlibro}, in order to characterize the integrability by quadratures of an involutive distribution of vector fields $\mathcal{Z}$ on a $n$-dimensional manifold \cite{basarab,hartl1994solvable,sherring1992geometric,Barco2001}. Roughly speaking,  a solvable structure for a distribution $\mathcal{Z}$ of rank $r$ consists of a sequence of $n-r$ vector fields that gives rise to a chain of distributions such that each vector field in the structure is a symmetry of the previous distribution. 

Almost at the same time, \cinf-symmetries  were introduced as a generalization of the classical Lie symmetry method of reduction \cite{olver86,ovsiannikovlibro}  for ordinary differential equations (ODEs) \cite{muriel01ima1}. Since its introduction  \cinf-symmetries have been extended in multiple directions \cite{gaetamorando,gaeta2,Morando_mu_deformation, gaetatwisted,cicognareduction,CicognaHamiltonian,cicogna2013dynamical,cicogna2012generalization,levi2010,levinucci,murielolver,cicognagaeta_noether,adrianJPA2018,nadjafikhah2013variational,PaolaPfaffian,RuizServan2022variational} and are being extensively used \cite{bhuvaneswari2012application,AbdelKader2013147,guha2013lambda,Gun2013,Polat20161571,Jafari2021,Kozlov2015,MENDOZAkundu,MENDOZAburgers,Mohanasubha2019318,Orhan201969,adrianAMC2018,Ruiz2020,zhang2008}, allowing to solve equations that may even lack Lie point symmetries \cite{muriel01ima1,muriel03lie,Cimpoiasu201522}.  

The idea that allowed to extend  the notion of Lie point symmetry to \cinf-symmetry in the context of ODEs, has been adapted in \cite{pancinf-sym,pancinf-struct} for involutive distributions of vector fields. The condition for a vector field to be a \cinf-symmetry of a distribution is less restrictive than for a symmetry, which implies that in practice the \cinf-symmetries of a distribution are easier to find than its symmetries. When in the notion of solvable structure we let the elements be \cinf-symmetries, instead of symmetries, of the chains of distributions mentioned above, we get a  more general structure,  which has been called \cinf-structure in \cite{pancinf-sym}. The key point in this new theory is that once a \cinf-structure for an involutive distribution $\mathcal{Z}$  of corank $k$  has been determined, then  $\mathcal{Z}$ can be integrated by  sequentially  solving $k$ integrable Pfaffian equations \cite[Theorem 3.5]{pancinf-sym}. These Pfaffian equations are defined in spaces whose dimensions decrease one unit at each stage. The Pfaffian equations are completely integrable, although unlike solvable structures, they may not be integrable by quadratures. The well-known result on the
relationship between integrating factors and Lie point symmetries for first-order ODEs \cite{olver86,stephani,blumanlibro,ibragimovlibro} has been recently  generalized in \cite{pancinf-struct} for \cinf-structures and involutive distributions of arbitrary corank by introducing symmetrizing factors.
Relevant results on the role of these symmetrizing factors on the integrability by quadratures of the Pfaffian equations
that arise by application of the \cinf-structure method have been also derived \cite{pancinf-struct}.

In this work we present some new applications of the integration procedure associated to \cinf-structures. The paper is organized as follows: in Sections \ref{prelim} and \ref{mainsection} we recall the main definitions and results in the theory of \cinf-structures, by adapting some of  the theoretical results that were obtained in \cite{pancinf-sym} to the problems that we address in this paper. 
In Section \ref{sec4} we explore the application of the \cinf-structure method to fully integrate two systems of first-order ordinary differential equations, one of which is a Lotka-Volterra system, frequently employed to describe the dynamics of biological systems. Additionally, we investigate three scalar ODEs in Section \ref{sec5}, two of which are of fourth order and one of third order. Notably, the considered equations exhibit a lack of sufficient Lie point symmetries, and even powerful symbolic systems like Maple fail to provide explicit solutions for them. Nevertheless, utilizing our novel integration method based on \cinf-structures, we achieve complete integration of the equations that are difficult to solve by using conventional methods.

\section{Preliminaries}\label{prelim}

In this paper functions, vector fields and  differential forms are assumed to be smooth (meaning $\mathcal{C}^{\infty}$) on a contractible open subset $U$ of $\mathbb{R}^n.$ 
In what follows, $\mathfrak{X}(U)$ and $\Omega^k(U)$ will denote the $\mathcal{C}^{\infty}(U)$-module of all smooth vectors and $k$-forms, respectively, while $\Omega^*(U)$ will denote the exterior differential algebra of all differential forms on $U.$  

Given a set  $\{Z_1,\ldots,Z_r\}$  of pointwise linearly independent vector fields on $U,$ by  $\mathcal{Z}:=\mathcal{S}(\{Z_1,\ldots,Z_r\})$  we denote  the  submodule  of $\mathfrak{X}(U)$  generated by $\{Z_1,\ldots,Z_r\}.$  Similarly, the submodule of $\Omega^1(U)$ generated by  a set of pointwise linearly independent 1-forms  $\{\sigma_1,\ldots,\sigma_s\}$ will be denoted by   $\mathcal{P}:=\mathcal{S}(\{\sigma_1,\ldots,\sigma_s\}).$  The submodule $\mathcal{Z}$ (resp. $\mathcal{P}$) defines a distribution (resp. a Pfaffian system) of constant rank $n-r$ (resp. $n-s$).

The annihilator of $\mathcal{Z}$ is the set of the differential forms $\omega \in \Omega^*(U)$ such that \newline $\omega(Y_1,\ldots,Y_{k})=0$ whenever $Y_1,\ldots, Y_k\in \mathcal{Z}.$ This set, which will be denoted  by $\mbox{Ann}(\mathcal{Z}),$ is an ideal of $\Omega^*(U)$ locally generated by $n-r$ pointwise linearly independent 1-forms $\{\omega_1,\ldots,\omega_{n-r}\}$  \cite{warner,bryant2013exterior}. In this case, we will write  $\mathcal{Z}^{\circ}=\mathcal{S}(\{\omega_1,\ldots,\omega_{n-r}\}).$ It can be checked that the Pfaffian system  $\mathcal{Z}^{\circ}$ can be characterized in terms of the interior product (or contraction) $\lrcorner$ as follows: 
$$ \mathcal{Z}^{\circ} = \{\omega\in \Omega^1(U):Z\,\lrcorner\,\omega=0, \,\mbox{for each} \, Z\in \mathcal{Z}\}.$$ 

Let us recall that the distribution $\mathcal{Z}$ is said to be involutive if $[Z_i,Z_j]\in\mathcal{Z}$ for $1\leq i,j\leq r.$  A well-known result  states that $\mathcal{Z}$ is involutive if and only if the ideal $\mbox{Ann}(\mathcal{Z})$ is closed under exterior differentiation $d,$ i.e., if  $\mbox{Ann}(\mathcal{Z})$ is a differential ideal (see, for instance, Proposition 2.30 and Definition 2.29 in \cite{warner}). 
In this case, Frobenius Theorem \cite[Theorem 1.60]{warner} guarantees, for each $p\in U,$  the local existence of a unique connected integral manifold of $\mathcal{Z}$ of maximal dimension \cite[Definition 1.63]{warner}. Such integral manifolds can  be  defined (locally) by the level sets of a complete set of first integrals $I_1,\ldots,I_{n-r}$ for the distribution $\mathcal{Z}.$  It is clear that, in this case, the independent 1-forms $\{dI_1,\ldots, dI_{n-r}\}$ generate the corresponding Pfaffian system $\mathcal{Z}^{\circ},$ which is said to be completely integrable  \cite{warner,bryant2013exterior}. In this sense, integrating a completely integrable Pfaffian system is equivalent to integrating the corresponding involutive system of vector fields. 

In such integration procedures, the notion of solvable structure,  introduced by Basarab-Horwath in \cite{basarab} plays a fundamental role (see also \cite{sherring1992geometric}). This concept is 
based on the notion of symmetry of a distribution, which generalizes Lie point symmetries: \cite{lychagin1991,basarab,lychagin2007contact}: 
\begin{definition}\label{def_sym}
     A symmetry of an involutive distribution $\mathcal{Z}$ is a vector field $X$ such that the set  $\{Z_1,\ldots,Z_r,X\}$ is  pointwise linearly independent on $U$ and $[X,\mathcal{Z}]\subset \mathcal{Z}$.
\end{definition} Now we can recall the concept of solvable structure: 
\begin{definition}\label{def_sol}\cite[Definiton 4]{basarab}:
  A solvable structure for $\mathcal{Z}$ consists of an ordered set{\footnote{Observe that in references  such as \cite{basarab,hartl1994solvable,sherring1992geometric,Barco2001,Barco2002,conmorando} the term solvable structure  refers to the whole ordered set $\langle Z_1,\ldots, Z_r, X_1,\ldots,X_{n-r}\rangle$ that also includes the vectors fields  that generate the distribution $\mathcal{Z}.$} } of vector fields $\langle X_1,\ldots,X_{n-r}\rangle$ such that $X_1$ is a symmetry of $\mathcal{Z}$ and $X_i$ is a symmetry of the distribution $\mathcal{Z}\oplus \mathcal{S}(\{X_1,\ldots,X_{i-1}\})$ for $i=2,\ldots,n-r.$ 
\end{definition}

The main result concerning solvable structures is that the knowledge of a solvable structure allows us to find the integral manifolds of $\mathcal{Z}$, at least locally, by quadratures alone \cite[Proposition 3]{basarab}. 
A dual version of Definition \ref{def_sol}, given in terms of differential 1-forms, was introduced in \cite[Defintion 4]{hartl1994solvable} by Hartl and Athorne. These authors also re-established the integrability result by Basarab-Horwath from a dual point of view (see \cite[Proposition 5]{hartl1994solvable}).  We refer the reader also to \cite{Barco2001,Barco2002}  for further details on the integration procedure  associated to solvable structures.

Solvable structures are very useful in the study of ordinary differential equations (ODEs), because such problems can be reformulated as the task of integrating  systems of vector fields or 1-forms. For instance, consider a system of first-order ODEs 
\begin{equation}\label{eq_dynsys}
\left\{\begin{array}{lcl}
\dot{x}_1&=&\phi_1(t,x_1,\ldots,x_n),\\
\dot{x}_2&=&\phi_2(t,x_1,\ldots,x_n),\\
&\vdots&\\
\dot{x}_n&=&\phi_n(t,x_1,\ldots,x_n),\\
\end{array}\right.
\end{equation}
where $\phi_1,\ldots,\phi_n$ are smooth functions on some open set $U\subset  \mathbb{R}^{n+1}$ and over dot  denotes differentiation with respect to the independent variable $t.$ Any solution of system \eqref{eq_dynsys} defines a one-dimensional integral manifold of the (trivially involutive) rank 1 distribution generated by the vector field
\begin{equation}\label{eq_campodynsys}
Z=\partial_t+\phi_1(t,x_1,\ldots,x_n) \partial_{x_1}+\ldots+\phi_n(t,x_1,\ldots,x_n)\partial_{x_n}.
\end{equation}

The extension to systems of ODEs of higher order is straightforward. Consider, for instance, a general $m$th-order ODE: 
\begin{equation}\label{eq_ode}
u_m=\phi(x,u^{(m-1)}),
\end{equation}
where $u^{(m-1)}=(u, u_1, \ldots, u_{m-1})$ denotes the dependent variable $u$ and,  for $1\leq k\leq m,$  $u_k$ denotes the derivative  of order $k$  of $u$ with  respect to the independent variable $x.$  By setting $x=t,$  $x_1=u,$ and $x_k=u_{k-1},$ for $1\leq k\leq m,$ then  equation (\ref{eq_ode}) can be transformed into a system of the form (\ref{eq_dynsys}), whose associated vector field (\ref{eq_campodynsys}), written in terms of original variables $(x,u^{(m-1)}),$ becomes
\begin{equation}\label{eq_campoode}
Z=\partial_x+u_1 \partial_{u}+\ldots+\phi(x,u^{(m-1)})\partial_{u_{m-1}}.
\end{equation} 
In this case, any integral manifold of the distribution generated by the vector field \eqref{eq_campoode} corresponds to the $(m-1)$th-prolongation of a solution of the equation \eqref{eq_ode} \cite{olver86,stephani,blumanlibro}.  

Therefore, the method of solvable structures can be applied to integrate the given ODE (or the system of ODEs) by quadratures alone. This outcome extends the classical result which states that an $n$th order system of $m$ differential equations, with a $mn$th-dimensional solvable algebra of Lie point symmetries, can be integrated by quadratures.  We refer  the reader to  \cite[Proposition 6]{hartl1994solvable} and \cite[Section V]{sherring1992geometric} for further details on the application of solvable structures to the integration of differential equations.

\section{\texorpdfstring{\cinf}--structures and integrability of distributions}\label{mainsection}

This notion of \cinf-symmetry for a distribution was introduced in \cite[Definition 3.2.]{pancinf-sym}, as a generalization of the idea of \cinf-symmetry for ODEs \cite{muriel01ima1}: 

\begin{definition}\label{Csymmdistribution}
A \cinf-symmetry of an involutive distribution $\mathcal{Z}=\mathcal{S}(\{Z_1,\ldots,Z_r\})$ is a vector field $X$ such that the set $\{Z_1,\ldots,Z_r,X\}$ is pointwise linearly independent on $U$ and the distribution $\mathcal{S}(\{Z_1,\ldots,Z_r,X\})$ is involutive.
\end{definition}

Note that by Definition \ref{def_sym} every symmetry $X$ of an involutive distribution $\mathcal{Z}$ is also a \cinf-symmetry of  $\mathcal{Z}$.

The previous notion of \cinf-symmetry of a distribution was used in \cite{pancinf-sym} to extend the concept of solvable structure as follows:

\begin{definition}\label{Qsolvable}\cite[Definition 3.3.]{pancinf-sym}
Let $\mathcal{Z}$ be an involutive distribution on $U$. An ordered set of vector fields $\langle X_1,\ldots, X_{n-r}\rangle$ is  a \cinf-structure for $\mathcal{Z}$ if $X_1$ is a \cinf-symmetry of $\mathcal{Z}$ and, for  $i=2,\ldots,n-r,$ $X_i$ is a \cinf-symmetry of the distribution $\mathcal{Z}\oplus \mathcal{S}(\{X_1,\ldots,X_{i-1}\}).$
\end{definition}

Observe that a  solvable structure for $\mathcal{Z}$ is a particular case of a \cinf-structure for $\mathcal{Z}$ where each $X_i$  a symmetry of $\mathcal{Z}\oplus \mathcal{S}(\{X_1,\ldots,X_{i-1}\})$ instead of a  \cinf-symmetry.

The main result concerning  \cinf-structures is that they can be used to integrate the distribution $\mathcal{Z}$ by solving successively $n-r$ completely integrable Pfaffian equations. Unlike solvable structures, such Pfaffian equations may not be integrable by quadratures:
\begin{theorem}\label{maintheorem}\cite[Theorem 3.5]{pancinf-sym}
Let $\mathcal{Z}$ be an involutive distribution on $U\subset \mathbb{R}^n.$ Any $\mathcal{C}^{\infty}$-structure for $\mathcal{Z}$ can be used to find the integral manifolds of $\mathcal{Z}$ by solving successively $n-r$ completely integrable Pfaffian equations.
\end{theorem}

The next subsection outlines a procedure that can be employed to integrate the distribution $\mathcal{Z}$ when we have  a \cinf-structure of vector fields. This procedure will be used in subsequent sections to integrate various distributions that emerge in problems modelled by differential equations.  

\subsection{\texorpdfstring{\cinf}--structure-based method of integration}\label{procedimiento}

Given a \cinf-structure of vector fields $\langle X_1,\dots,X_r\rangle$ for $\mathcal{Z},$ a method that can be used to integrate $\mathcal{Z}$ by applying Theorem \ref{maintheorem} proceeds as follows. Let $(x_1,\dots,x_n)$ be a system of local coordinates on  $U \subset \mathbb{R}^n$ and consider  the volume form $\boldsymbol{\Omega} = dx_1 \wedge \cdots \wedge dx_n$.  

Let us introduce the 1-forms 
\begin{equation}\label{formas}
    \omega_i={X_{n-r}\,\lrcorner\,\ldots\,\lrcorner\,\widehat{X_i}\,\lrcorner\,\ldots\,\lrcorner\,X_{1}\,\lrcorner\,Z_r \contract \ldots \contract Z_1\,\lrcorner\,\boldsymbol{\Omega}}, \quad 1\leq i\leq n-r,
    \end{equation}
    where $\widehat{X_i}$ indicates omission of $X_i,$ and set  \begin{equation}\label{pfaff}\mathcal{P}_i:=\mathcal{S}(\{\omega_{i+1},\ldots,\omega_{n-r}\}), \quad 0\leq i\leq n-r-1. \end{equation} It follows from \eqref{formas} that 
\begin{equation}
\label{dual}
 \begin{array}{l} \mathcal{P}_0=\mathcal{Z}^{\circ},\\
 \mathcal{P}_i=\left(\mathcal{Z}\oplus \mathcal{S}(\{X_1,\ldots,X_i\})\right)^{\circ},\quad 1\leq i\leq n-r-1.
\end{array} \end{equation}

Considering that the distribution $\mathcal{Z}$ is involutive and that, according to Definition \ref{Qsolvable}, the distributions $\mathcal{Z}\oplus \mathcal{S}(\{X_1,\ldots,X_{i}\})$  for   $1\leq i \leq n-r-1$ are also involutive, then it can be deduced from   \eqref{dual} that the  Pfaffian systems given in \eqref{pfaff} are completely integrable. More explicitly,  there exist 1-forms $\sigma_i$ for $1\leq i\leq n-r$ such that
\begin{equation}\label{d_cerrada}
d\omega_i=\sigma_i\wedge\omega_i+\sum_{j=i+1}^{n-r}\sigma_{i}^j\wedge \omega_j, 
\end{equation}
for some 1-forms $\sigma_{i}^j,$ $j=i+1,\ldots,n-r.$


Since the integration of the involutive distribution $\mathcal{Z}$ is equivalent to the integration of the Pfaffian system $\mathcal{P}_0,$ we describe below how to integrate $\mathcal{P}_0$ step by step: 
\begin{enumerate}
    \item For $i=n-r,$ equation \eqref{d_cerrada} becomes  $d\omega_{n-r}= \sigma_{n-r}\wedge\omega_{n-r},$ which implies that the Pfaffian equation  $\omega_{n-r}\equiv 0$ is Frobenius integrable. A first integral $I_{n-r}=I_{n-r}(x_1,\ldots, x_n)$ for $\mathcal{P}_{n-r-1}$ is any particular solution to  the system of linear  first-order PDEs arising from the condition
    $$
    dI_{n-r}\wedge \omega_{n-r}=0.
    $$  
    For $C_{n-r}\in \mathbb{R},$ the level set of $I_{n-r}$
    \begin{equation}
        \Sigma_{(C_{n-r})}=\{x\in U \subset\mathbb{R}^n: I_{n-r}(x)=C_{n-r}\}
    \end{equation} defines an integral submanifold, of dimension $n-1,$ of the distribution \newline $\mathcal{Z}\oplus \mathcal{S}(\{X_1,\ldots,X_{n-r-1}\}).$
   
    \item For $1\leq i\leq n-r,$ we denote by $\omega_i|_{\Sigma_{(C_{n-r})}}$ the restriction of  $\omega_i$ to $\Sigma_{(C_{n-r})}.$ Observe that $\omega_{n-r}|_{\Sigma_{(C_{n-r})}}=0.$   
    The restriction to $\Sigma_{(C_{n-r})}$ of equations \eqref{d_cerrada} for $1\leq i\leq i=n-r-1,$ implies that 
 $\omega_{n-r-1}|_{\Sigma_{(C_{n-r})}}$ is Frobenius integrable.     As before, a corresponding first integral $I_{n-r-1}=I_{n-r-1}(x;C_{n-r}),$ defined for $x$ in some open set of  $\Sigma_{(C_{n-r})},$ is given by any particular solution to  the system of linear homogeneous  first-order PDEs arising from the condition
$$
dI_{n-r-1}\wedge \omega_{n-r-1}|_{\Sigma_{(C_{n-r})}}=0.
$$  
For $C_{n-r-1}\in \mathbb{R},$ the submanifold of $\Sigma_{(C_{n-r})}$ defined by the level set $I_{n-r-1}=C_{n-r-1}$ is an integral manifold of the Pfaffian equation $\omega_{n-r-1}|_{\Sigma_{(C_{n-r})}}\equiv 0$,  that will be denoted by $\Sigma_{(C_{n-r-1},C_{n-r})}.$

 \item   We continue this process,  taking into account that in each stage we integrate a 1-form defined in a space whose dimension is one unit lower than in the previous step. At the end, we obtain the integral manifold $\Sigma_{(C_{1},\ldots,C_{n-r})}$   of $\mathcal{Z}^{\circ},$  expressed  in implicit form as $I_{1}=C_{1},$ where  $I_1$ denotes the first integral that arises after integrating the last Pfaffian equation $\omega_{1}|_{\Sigma_{(C_2,\ldots, C_{n-r})}}\equiv 0.$ 
\end{enumerate}

\vspace{0.5cm}
The theoretical foundation behind the procedure above is explained in \cite[Theorem 3.5]{pancinf-sym}. Readers interested in a closer exploration of the \cinf-structure integration process and in reviewing illustrative examples are referred to Sections 3.3 and 3.4 in \cite{pancinf-struct}. 

In addition, if an element $X_i$ of the \cinf-structure is not only a \cinf-symmetry of $\mathcal{Z}\oplus \mathcal{S}(\{X_1,\ldots,X_{i-1}\})$ but a symmetry, then the corresponding Pfaffian equation at the $i$th stage can be solved by quadrature using a (relative) integrating factor (see Theorem 4.1 and Remark 4.3 in \cite{pancinf-struct} for details). The integrability of the distribution by quadrature via solvable structures turns out to be a special case of the more general \cinf-structure integration method.

In the following sections we use the  integration method described above to find  exact solutions to several problems modelled by ordinary differential equations.

\section{\texorpdfstring{\cinf}--structures for systems of first-order ODEs}\label{sec4}

In this section we examine the application of the \cinf-structure method to systems of first-order ODEs. 

The first system describes a Lotka-Volterra model previously considered by P. Basarab-Horwarth in his paper on solvable structures \cite{basarab}. His procedure requires three vector fields to produce two independent first integrals of the system. In the following subsection we show that only one of these vector fields is needed to construct a \cinf-structure which can be used to completely solve the system.

\subsection{A Lotka-Volterra model}
Lotka-Volterra models, also known as predator-prey models, are systems of first order ODEs used to describe the dynamics between two or more interacting species in an ecosystem, typically a predator and its prey.
The Lotka-Volterra model is a simple but powerful tool for understanding the dynamics of predator-prey interactions and has applications in fields such as ecology, biology, and economics (see, for example, \cite{takeuchi1996global,GRAMMATICOS1990683,solomon2000generalized,maier2013integration} for further details).

P. Basarab-Horwath in \cite[Section 4]{basarab} applied a method based on  solvable structures to find two first integrals for a biparametric family of 3D Lotka-Volterra models
\begin{equation}\label{c3eqLV}
\left\{
\begin{array}{l}
\dot{x}(t)=zx-\dfrac{yx}{AB},\\[7px]

\dot{y}(t)=xy+Azy,\\[7px]

\dot{z}(t)=Bxz+yz,\\
\end{array}
\right.
\end{equation}
with arbitrary constants $A, B\in \mathbb R$, $A,B \neq 0$. 
More specifically, he provided two vector fields
\begin{equation}\label{Ys}
    \begin{array}{lll}
    Y_1  & =&z\partial_x+Ay\partial_y+(y+Bz)\partial_z,\\
    Y_2  & =& x\partial_x-ABz\partial_z.
    \end{array}
\end{equation}
which are in involution with the vector field corresponding to the system:
$$
Z=\left(zx-\dfrac{yx}{AB}\right)\partial_x+ \left(xy+Azy\right)\partial_y+\left(Bxz+yz\right)\partial_z,
$$ 
as it can be checked through the corresponding commutation relationships. However, neither  $\langle Y_1,Y_2\rangle$
nor $\langle Y_2,Y_1\rangle$ constitutes a solvable structure for $\mathcal{S}(\{Z\}),$ because $[Y_1,Y_2]\notin \mathcal{S}(\{Z,Y_1\})$ and $[Y_1,Y_2]\notin \mathcal{S}(\{Z,Y_2\}).$ For this reason, P. Basarab-Horwath had to provide an additional vector field 
\begin{equation}\label{sym}
V=x\partial_x+y\partial_y+z\partial_z,   
\end{equation}
which is a symmetry of $\mathcal{S}({Z})$ and commutes with $Y_1$ and $Y_2.$ This implies that $V$ is a symmetry of both involutive distributions $\mathcal{S}(\{Z,Y_1\})$ and $\mathcal{S}(\{Z,Y_2\})$. Applying the theoretical results on solvable structures, the symmetry $V$ was used in \cite{basarab} to integrate, separately  and by quadratures, the distributions $\mathcal{S}(\{Z,Y_1\})$ and $\mathcal{S}(\{Z,Y_2\})$.  

A first integral for $\mathcal{S}(\{Z,Y_1\})$ is
\begin{equation}\label{phi1}
    \varphi_1=ABx+y-Az,
\end{equation} while
\begin{equation}\label{phi2}
    \varphi_2=x^{AB}y^{-B}z
\end{equation}
is a first integral for $\mathcal{S}(\{Z,Y_2\})$. These first integrals   are functionally independent, because $Z\wedge Y_1\wedge Y_2\neq 0.$

It is interesting to note that only one of the  vector fields $Y_1$ or $Y_2$  is necessary to integrate system \eqref{c3eqLV} by the \cinf-structure method: since $\mathcal{S}(\{Z,Y_1\})$ is an involutive distribution, $Y_1$ can be chosen as the first element of a \cinf-structure for $\mathcal{S}(\{Z\})$. The last element can be any vector field independent with $\{Z,Y_1\},$ such as $\partial_z.$ Therefore, $\langle Y_1, \partial_z\rangle$  defines a   \cinf-structure for $\mathcal{S}(\{Z\})$ and it can be used to integrate the system by the procedure described in Section \ref{procedimiento}. 
The same procedure could be followed by using $Y_2$ instead $Y_1,$ because  $\langle Y_2, \partial_z\rangle$ is also a \cinf-structure for $\mathcal{S}(\{Z\})$.  

Nevertheless, instead of using one of these two \cinf-structures, which require the knowledge of at least one of the vector fields $Y_1$ or $Y_2$, we show how to construct a \cinf-structure for $\mathcal{S}(\{Z\})$ directly, without using the vector fields provided by Basarab-Horwath. It is worth noting that the method used to obtain these vector fields has not been explained in \cite{basarab}.

In order to find a \cinf-structure for $\mathcal{S}(\{Z\}),$ we first observe that a if vector field $X_1$ is a \cinf-symmetry of  $\mathcal{S}(\{Z\}),$ then so is any vector field in $\mathcal{S}(\{Z,X_1\}).$ 
This allows us to simplify the search for $X_1$ by assuming that its form is
$$
X_1=\partial_y+g(x,y,z) \partial_z.
$$   

According to Definition \ref{Qsolvable}, $X_1$ must satisfy  the condition $[X_1,Z] \in \mathcal{S}(\{Z,X_1\}).$ Equivalently, the 1-form
$\omega_2=X_1\contract Z\contract \boldsymbol{\Omega},$ where $\boldsymbol{\Omega}=dx\wedge dy\wedge dz,$ satisfies $\omega_2\wedge d\omega_2=0,$ i.e., the Pfaffian equation $\omega_2\equiv 0$ is  completely integrable. Any of these two equivalent conditions yields a determining equation for the function $g=g(x,y,z).$ It can be checked that such PDE is of the form 
\begin{equation}\label{gecudet}
    \rho_1\,g_x+ \rho_2\,g_y+ \rho_3\,g_z+ (Ag-1) (\rho_4\,g+\rho_5)=0,
\end{equation} 
where we omit the explicit  expressions of the functions $\rho_i=\rho_i(x,y,z),$ for $1\leq i\leq 5,$ because they are irrelevant for the following discussion. A  particular solution of the determining equation \eqref{gecudet} arises immediately, the constant function 
\begin{equation}
    \label{sol}
    g(x,y,z)=\dfrac{1}{A}.
\end{equation}
Therefore, the vector field 
\begin{equation}\label{X1}
X_1=\partial_y+\frac{1}{A} \partial_z
\end{equation}
is a \cinf-symmetry of $\mathcal{S}(\{Z\})$. As the second element of the \cinf-structure we can choose any vector field $X_2$ such that $\{Z,X_1,X_2\}$ are linearly independent. For instance, we can use the vector field $X_2=\partial_z$.

Once the \cinf-structure $\langle X_1,X_2\rangle$ for $\mathcal{S}(\{Z\})$ has been determined, we calculate the 1-forms $\omega_1$ and $\omega_2$ given in \eqref{formas}:
\begin{equation}\label{w1w2}
    \begin{array}{rl}
\omega_1=X_2\contract Z\contract \boldsymbol{\Omega}&=\left(Ayz+xy\right)dx-\dfrac{x\left(ABz-y\right)}{AB}dy,\\[10pt]
\omega_2=X_1\contract Z\contract \boldsymbol{\Omega}&= 
      -\dfrac{x\left(ABz-y\right)}{A^2B} \left(ABdx+dy-Adz\right).
\end{array}
\end{equation}

The Pfaffian equation $\omega_2\equiv 0$ is completely integrable and a corresponding first integral $I_2=I_2(x,y,z)$ arises from the condition $dI_2\wedge \omega_2=0,$ which yields the following system of PDEs:
\begin{equation}\label{edet}
    \begin{array}{lll}
        A (I_2)_y+ (I_2)_z& =& 0,\\
          (I_2)_x+ B (I_2)_z& =& 0. 
    \end{array}
\end{equation}
The first equation implies that $I_2=F\left(x,r\right),$ where $r=Az-y$ and $F=F(x,r)$ is, in principle, an arbitrary smooth function. Then  the second equation in \eqref{edet} becomes
$$F_x+AB F_r=0,$$ from which the particular solution $F(x,r)=ABx-r$ arises immediately. Therefore, a   first integral for $\omega_2\equiv 0$ is given by $I_2= F\left(x,Az-y\right):$
\begin{equation}\label{I2}
    I_2=ABx-Az+y.
\end{equation}

Observe that $I_2=\varphi_1,$ where  $\varphi_1$ is the first integral \eqref{phi1} provided by Basarab-Horwath.
In order to find the remaining first integral, we restrict $\omega_1$ to the submanifold  $\Sigma_{(C_2)}$ implicitly defined by $I_2=C_2,$ where $C_2\in \mathbb{R}:$ 
$$
\omega_1|_{\Sigma_{(C_2)}}=\left(ABxy-C_2y+xy+y^2\right)dx-\frac{x\left(AB^2x-BC_2+By-y\right)}{AB}dy.
$$
The Pfaffian equation $\omega_1|_{\Sigma_{(C_2)}}\equiv 0$ is completely integrable. 
It can be checked that $$\mu=\dfrac{1}{xy(ABx+y-C_2)}$$ 
is an integrating factor for $\omega_1|_{\Sigma_{(C_2)}}.$ A corresponding primitive $I_1=I_1(x,y;C_2)$ arises after integrating two  rational functions: 
\begin{equation}\label{I1volt}
    {I}_1=\ln(|x^{AB}y^{-B}(ABx-C_2+y)|).
\end{equation} 
If  $C_2$ in \eqref{I1volt} is replaced by the right-hand side of \eqref{I2} we get the function $J_1(x,y,z)=I_1(x,y;I_2):$
$$
J_1=Ax^{AB}y^{-B}z,
$$
which, up to a constant, coincides with the first integral $\varphi_2$ in \eqref{phi2},  previously obtained in  \cite{basarab}. 

The orbits of the system \eqref{c3eqLV} can be expressed in implicit form as follows: 
\begin{equation}\label{or:lt}
    Ax^{AB}y^{-B}z=C_1,\quad ABx-Az+y=C_2,\quad (C_1,C_2\in \mathbb{R}).
\end{equation}

In consequence, the \cinf-structure method provides an alternative approach to integrate system \eqref{c3eqLV}. In this procedure, only the vector field \eqref{X1} has been used, instead of the three vector fields $Y_1,Y_2$ and $V$ in \eqref{Ys} and \eqref{sym} required in   \cite{basarab} by using solvable structures techniques.

\subsection{Integration of a non-autonomous system through \texorpdfstring{\cinf}--structures}
In the following example we study a system of first order ODEs which, to our knowledge, can not be easily solved by classical procedures. 
We will show how to construct  a \cinf-structure for the system and how to use it to find its general solution, expressed in terms of a complete set of solutions of a linear second-order homogeneous equation. 

Consider the system of first-order ODEs:
\begin{equation}\label{c3eqSist}
\left\{\begin{array}{l}
\dot{x}(t)=\dfrac{ 2 t y-t^2 x^2-2 y x^2-x^2 }{2 tx},\\[7px]
\dot{y}(t)=t-x^2 y,
\end{array}\right.
\end{equation}
with associated vector field
$$
Z=\partial_t+\dfrac{ 2 t y-t^2 x^2-2 y x^2-x^2 }{2 tx}\partial_x+(t-x^2 y)\partial_y,
$$
defined on the open set
\begin{equation}\label{Mej4.2}
    M=\{(t,x,y)\in \mathbb{R}^3: tx\neq 0\}.
\end{equation}

To find the first element $X_1$ of a \cinf-structure for the distribution $\mathcal{S}(\{Z\})$ we assume, as in the previous example, that $X_1$ is of the form 
$X_1=\partial_x+g(t,x,y)\partial_y.$ The determining equation for the function $g(t,x,y)$ can be obtained from the condition $[X_1,Z]\in \mathcal{S}(\{Z,X_1\}).$ This is equivalent to the condition $\omega_2\wedge d\omega_2=0,$ where $\omega_2=X_1\contract Z\contract\boldsymbol{\Omega}$ for $\boldsymbol{\Omega}=dt\wedge dx\wedge dy.$  

In order to ease the search of a particular solution of this determining equation, we can begin by trying to find a particular solution of the form $g(t,x,y)=f(t)h(x).$  It can be checked that by canceling out the coefficients of $y$ we obtain a  system of determining equations for the functions $f=f(t)$ and $h=h(x)$ that, after some calculations, becomes
$$
h(x)f(t)=  tx,\quad
    f'(t) = \dfrac{f(t)}{t}.
$$ 
By choosing the  particular solution  
$$
h(x)=x,\quad f(t)=t,
$$ 
we obtain that the vector field $X_1=\partial_x+tx\partial_y$ is a \cinf-symmetry of the distribution $\mathcal{S}(\{Z\})$ and hence it can be selected as the first element of a \cinf-structure for  $\mathcal{S}(\{Z\}).$ As a second element  we can choose any vector field $X_2$ such that the set $\{Z,X_1,X_2\}$ is linearly independent, such as, for instance, $X_2=\partial_y.$  Therefore the vector fields 
     \begin{equation}\label{csym_sistema2}
        X_1=\partial_x+tx\partial_y, \quad X_2=\partial_y
    \end{equation} constitute a \cinf-structure for $\mathcal{S}(\{Z\}).$ The corresponding 
    commutations relationships become
\begin{align}      \label{corchetes_sistema}
&[X_1,Z] = \frac{-2tx^4+t^2x^2-2x^2y-2ty-x^2}{tx^2}\, X_1, \\
&[X_2,Z] = \frac{-x^2+t}{tx}\,X_1 -t\,X_2, \\
&[X_2,X_1] = 0.
\end{align}

It is crucial to emphasize that neither $X_1$ is a symmetry of $\mathcal{S}({Z})$, nor $X_2$ is a symmetry of $\mathcal{S}({Z,X_1})$. Specifically, $X_1$ and $X_2$ do not correspond to symmetries of the system \eqref{c3eqSist}. As a result, the integration method based on the \cinf-structure presented here provides a novel alternative to conventional symmetry procedures.

The integration procedure by using the \cinf-structure defined by \eqref{csym_sistema2} proceeds as follows: the corresponding 1-forms given in \eqref{formas} become
\begin{align}
\label{w1}& \omega_1=\dfrac{2 t y-t^2 x^2-2 x^2 y-x^2}{2 x t} dt- dx,\\
\label{w2}& \omega_2=\left(ty -\frac{1}{2}t^2 x^2-\frac{1}{2}x^2-t\right) dt-tx dx+dy.
\end{align}
The Pfaffian equation $\omega_2\equiv 0$ is completely integrable;  it can be checked that a corresponding first integral is given by the smooth function
\begin{equation}\label{I2_sistema2}
    I_2(t,x,y)=\frac{1}{2}e^{\frac{1}{2}t^2}(tx^2-2 y+2).
\end{equation}
The restriction of the 1-form $\omega_1$ given in \eqref{w1} to the level set $I_2=C_2$,  $C_2\in \mathbb{R},$ denoted by ${\Sigma_{(C_2)}},$ becomes
\begin{equation}\label{w1res}
\omega_1|_{\Sigma_{(C_2)}}=\dfrac{2 t-tx^4-3x^2+2C_2(t-x^2) e^{-\frac{1}{2}t^2}}{2tx}dt-dx.
\end{equation}
In order to solve the Pfaffian equation $\omega_1|_{\Sigma_{(C_2)}}\equiv 0,$  we introduce the change $\bar{x}=x^2$ which transforms the ODE associated to the Pfaffian equation into the Riccati-type equation
\begin{equation}\label{ricati}
    \bar{x}'(t)=- \bar{x}(t)^2-\dfrac{2 C_2e^{-\frac{1}{2}t^2}+3}{t}\bar{x}(t)+2 (C_2e^{-\frac{1}{2}t^2}+1).
\end{equation} The standard transformation $\bar{x}(t)=\psi'(t)/\psi(t)$ transforms the Riccati-type equation  \eqref{ricati} into the following  linear second-order  homogeneous ODE:
\begin{equation}\label{lineal}
 \psi''(t)+ \left(\dfrac{2 C_2e^{-\frac{1}{2}t^2}+3}{t}\right) \psi'(t)-2(C_2e^{-\frac{1}{2}t^2}+1)\psi(t)=0.
\end{equation}
Let $\psi_1=\psi_1(t;C_2)$ and $\psi_2=\psi_2(t;C_2)$ be a fundamental set of solutions to the linear ODE \eqref{lineal}. These functions can be used to express  a first integral associated to the Riccati equation \eqref{ricati} (see, for instance, Proposition 4.1 in \cite{adrianJPA2017}). As a consequence, a first integral of the   Pfaffian equation defined by \eqref{w1res} becomes: 
\begin{equation}
    I_1(t,x;C_2)=\dfrac{-x^2\psi_1(t;C_2)+\psi_1'(t;C_2)}{-x^2\psi_2(t;C_2)+\psi_2'(t;C_2)}.
\end{equation}
By replacing $C_2$ by the right-hand side of \eqref{I2_sistema2} we get the function $J_1(t,x,y)=I_1(t,x;I_2)$, which is a first integral of $\mathcal{S}(\{Z\})$: 

\begin{equation}\label{J1}
    J_1(t,x,y)=\dfrac{-x^2\psi_1(t;\frac{1}{2}e^{\frac{1}{2}t^2}(tx^2-2 y+2))+\psi_1'(t;\frac{1}{2}e^{\frac{1}{2}t^2}(tx^2-2 y+2))}{-x^2\psi_2(t;\frac{1}{2}e^{\frac{1}{2}t^2}(tx^2-2 y+2))+\psi_2'(t;\frac{1}{2}e^{\frac{1}{2}t^2}(tx^2-2 y+2))}.
\end{equation}
From $I_1(t,x;C_2)=C_1$  and $I_2(t,x,y)=C_2$ where $C_1,C_2\in \mathbb{R},$ we get the general solution to system \eqref{c3eqSist}: 

\begin{equation}\label{x2}
\left\{\begin{array}{lll}
  x(t)   & =& \pm\left(\dfrac{C_1\psi_2'(t;C_2)-\psi_1'(t;C_2)}{C_1\psi_2(t;C_2)-\psi_1(t;C_2)}\right)^{1/2},\\
    y(t) &  =&1+C_2e^{-\frac{1}{2}t^2}+\dfrac{t}{2}\,\dfrac{C_1\psi_2'(t;C_2)-\psi_1'(t;C_2)}{C_1\psi_2(t;C_2)-\psi_1(t;C_2)}.
\end{array}\right.
\end{equation} 
where $\psi_1=\psi_1(t;C_2)$ and $\psi_2=\psi_2(t;C_2)$ are two functionally independent solutions to the linear ODE \eqref{lineal}. 

\subsubsection{Some particular families of solutions}

For particular values of the arbitrary constant $C_2\in \mathbb{R},$ the solutions to the corresponding linear ODE \eqref{lineal} are well-known special functions. For instance, for $C_2=0,$ equation  \eqref{lineal} becomes 
\begin{equation}\label{linealC20}
  t^2\psi''(t)+ 3t \psi'(t)-2t^2\psi(t)=0.
\end{equation}
Through the change of variables 
\begin{equation}\label{cambio}
    z=\sqrt{2}t,\quad \phi(z)=t\psi(t),
\end{equation}
equation \eqref{linealC20} becomes the modified Bessel equation
\begin{equation}\label{bessel}
  z^2\phi''(z)+ z \phi'(z)-(1+z^2)\phi(z)=0.
\end{equation} 
A fundamental set of solutions to equation \eqref{bessel} are   the modified Bessel functions $\mathbf{I}_{1}$ and $\mathbf{K}_{1}$ of the first and second kinds, respectively \cite{handbookfunctions}.  Therefore, according to \eqref{cambio}, the functions  
\begin{equation}\label{besselsol}
\psi_1(t)=\dfrac{1}{t} \mathbf{I}_{1}(\sqrt{2}t),\qquad \psi_2(t)=\dfrac{1}{t} \mathbf{K}_{1}(\sqrt{2}t),
\end{equation} 
are two linearly independent solutions to equation \eqref{linealC20}. As a consequence, a 1-parameter family of solutions to system \eqref{c3eqSist}, which corresponds to \eqref{x2} when $C_2=0,$ can be expressed in terms of the modified Bessel functions as follows: 
\begin{equation}\label{x2p}
\left\{\begin{array}{lll}
  x(t)   & =& \pm\left(\sqrt{2}\dfrac{C_1\mathbf{K}_{1}'(\sqrt{2}t)- \mathbf{I}_{1}'(\sqrt{2}t)}{C_1\mathbf{K}_{1}(\sqrt{2}t)-\mathbf{I}_{1}(\sqrt{2}t)}-\dfrac{1}{t}\right)^{1/2},\\
    y(t) &  =&\dfrac{1}{2}+\dfrac{\sqrt{2}t}{2}\,\dfrac{C_1\mathbf{K}_{1}'(\sqrt{2}t)- \mathbf{I}_{1}'(\sqrt{2}t)}{C_1\mathbf{K}_{1}(\sqrt{2}t)-\mathbf{I}_{1}(\sqrt{2}t)}.
\end{array}\right.
\end{equation} 

The derivatives of the modified Bessel functions $\mathbf{I}_{1}$ and $ \mathbf{K}_{1}$ can be expressed in terms of the modified Bessel functions $\mathbf{I}_{0}$ and $ \mathbf{K}_{0}$ \cite{handbookfunctions}:
$$\mathbf{K}_{1}'(z)=
\mathbf{K}_0(z)-\dfrac{1}{z}\mathbf{K}_1(z),\quad
\mathbf{I}_{1}'(z)=
\mathbf{I}_0(z)-\dfrac{1}{z}\mathbf{I}_1(z).
$$
Then
$$\mathbf{K}_{1}'(\sqrt{2}t)=-
\mathbf{K}_0(\sqrt{2}t)-\dfrac{\sqrt{2}}{2t}\mathbf{K}_1(\sqrt{2}t),\quad
\mathbf{I}_{1}'(\sqrt{2}t)=
\mathbf{I}_0(\sqrt{2}t)-\dfrac{\sqrt{2}}{2t}\mathbf{I}_1(\sqrt{2}t),
$$ 
and therefore \eqref{x2p} becomes
\begin{equation}
    \label{x2psinderivdas}
\left\{\begin{array}{lll}
  x(t)   & =& \pm\left(\sqrt{2}\dfrac{C_1
\mathbf{K}_0(\sqrt{2}t)-
\mathbf{I}_0(\sqrt{2}t)}{C_1\mathbf{K}_{1}(\sqrt{2}t)-\mathbf{I}_{1}(\sqrt{2}t)}-\dfrac{2}{t}\right)^{1/2},\\
    y(t) &  =&\dfrac{\sqrt{2}t}{2}\,\dfrac{C_1
\mathbf{K}_0(\sqrt{2}t)-
\mathbf{I}_0(\sqrt{2}t)}{C_1\mathbf{K}_{1}(\sqrt{2}t)-\mathbf{I}_{1}(\sqrt{2}t)}.
\end{array}\right.
\end{equation}

\section{\texorpdfstring{\cinf}--structures for scalar ODEs with a lack of Lie point symmetries}\label{sec5}

In this section we present a collection of  ordinary differential equations whose  symmetry algebras are either trivial or of lower dimension than the order of the ODE. In the latter scenario, the Lie method encounters certain obstacles when attempting to  obtain the general solution. However, we demonstrate how the \cinf-structures method successfully overcomes these difficulties and provides exact solutions of the equations under investigation. 

\subsection{A third-order ODE with two-dimensional algebra of Lie point symmetries}

In this example we consider a third-order ODE:
\begin{equation}\label{orden3}
     u_3=(u_1-u) u_2-\dfrac{u_1^2}{2}+u_1+\dfrac{u^2}{2},
\end{equation}
whose associated vector field is
$$
Z=\partial_x + u_1 \partial_u+u_2 \partial_{u_1}+\left((u_1-u) u_2-\frac{1}{2}(u_1^2-u^2)+u_1\right) \partial_{u_2}.
$$

It can be checked that the symmetry algebra of equation \eqref{orden3} is two dimensional and spanned by $\partial_x$ and $e^x\partial_u.$ By employing the Lie method of reduction, the transformation
\begin{equation}\label{transformacion}
z=u_1-u,\quad h(z)=u_2-u_1,\end{equation} leads to the  first-order ODE
\begin{equation}\label{abel}
h(z) h'(z)=(z-1)h(z)+\dfrac{z^2}{2}.
\end{equation}
Equation \eqref{abel} is an Abel-type equation whose general solution can be expressed in implicit form in terms of the modified Bessel functions of the first and second kinds $\mathbf{I}_{\alpha}$ and $\mathbf{K}_{\alpha},$ for $\alpha=0,1$ \cite{handbookfunctions}:
\begin{equation}\label{solabel}
    \dfrac{z\mathbf{K}_{0}\left(-\sqrt{z^2-{2}{h(z)}}\right)-\sqrt{z^2-{2}{h(z)}}\mathbf{K}_{1}\left(-\sqrt{z^2-{2}{h(z)}}\right)}{z\mathbf{I}_{0}\left(\sqrt{z^2-{2}{h(z)}}\right)-\sqrt{z^2-{2}{h(z)}}\mathbf{I}_{1}\left(\sqrt{z^2-{2}{h(z)}}\right)}=C_1.
\end{equation} 
The recovery of solutions to equation \eqref{orden3} from \eqref{solabel}, by means of the transformation \eqref{transformacion}, seems to be infeasible. 

For this reason, we intend to integrate equation \eqref{orden3} by using the \cinf-structures method. Similar to the previous examples, finding the  elements of a \cinf-structure can significantly be simplified by assuming some of the infinitesimals to be  constant or linear in $u_1$. By following this approach, we obtain the following independent vector fields
$$
\begin{array}{l}
X_1=\partial_u+\partial_{u_1}+\partial_{u_2},\\
X_2=\partial_{u_1}+(u_1-u+1)\partial_{u_2},\\
X_3=\partial_{u_2}.\\
\end{array}
$$
They form a \cinf-structure for $\mathcal{S}(\{Z\}),$ as can be verified using the Lie brackets:
$$
\begin{array}{l}
[X_1,Z]=X_1,\\

[X_2,Z]=X_1+(u_1-u) X_2,\\

[X_2,X_1]=0.\\
\end{array}
$$

It is important to emphasize that neither $X_1$ is a symmetry of $\mathcal{S}(\{Z\}),$ nor $X_2$ is a symmetry of $\mathcal{S}(\{Z,X_1\}).$ In particular, nor $X_1$ neither $X_2$ correspond to symmetries of equation \eqref{orden3}.

We use the volume form $\boldsymbol{\Omega}=dx\wedge du\wedge du_1\wedge du_2$ to construct the corresponding 1-forms given in \eqref{formas}:
\begin{equation}\label{omegas_ej5.1}
    \begin{array}{l}
\omega_1=-u_1 dx+du,\\
\omega_2= (u_2-u_1)dx+du-du_1,\\
\omega_3=(u_2+u u_1-u_1+\frac{1}{2}(u_1^2+u^2))dx+(u_1-u)du+(u-u_1-1)du_1+du_2.\\
\end{array}
\end{equation}

A first integral for the first Pfaffian equation $\mathcal{P}_2,$ i.e., a function $I_3=I_3(x,u,u_1,u_2)$ such that $dI_3 \wedge \omega_3=0$, is given by
$$
I_3=e^x\left(u_2-\frac{1}{2}(u_1-u)^2-u_1\right).
$$

Let $\Sigma_{(C_3)}$ denote, as before, the level set $I_3=C_3,$ for $C_3\in \mathbb R.$ The restriction of the 1-form $\omega_2$ in \eqref{omegas_ej5.1} to $\Sigma_{(C_3)}$  becomes
\begin{equation}\label{omega2res}
\omega_2|_{\Sigma_{(C_3)}}=\left(\frac{1}{2}(u_1-u)^2+C_3 e^{-x}\right)dx+du-du_1.
\end{equation}

In order to continue the integration process, we need to distinguish the following cases: 
\begin{enumerate}
\item {\bf Case I: $C_3>0.$}

It can be checked that a function $I_2=I_2(x,u,u_1;C_3)$ such that $dI_2\wedge \omega_2|_{\Sigma_{(C_3)}}=0$ becomes: 
$$
I_2=\dfrac{e^{x/2}(u_1-u)\mathbf{J}_0({\sqrt{2 C_3} e^{-x/2}})+\sqrt{2 C_3}\mathbf{J}_1(\sqrt{2 C_3} e^{-x/2})}{e^{x/2}(u-u_1)\mathbf{Y}_0(\sqrt{2 C_3} e^{-x/2})-\sqrt{2 C_3}\mathbf{Y}_1(\sqrt{2 C_3} e^{-x/2})},  
$$ 
where  $\mathbf{J}_{\alpha},\mathbf{Y}_{\alpha}$ are the Bessel functions  of first and second kind, respectively \cite{handbookfunctions}. 

Let $\Sigma_{(C_2,C_3)}$ denote the submanifold of $\Sigma_{(C_3)}$ defined by $I_2(x,u,u_1;C_3)=C_2,$ where $C_2\in \mathbb{R}.$ The restriction of the 1-form $\omega_3$ in \eqref{omegas_ej5.1} to $\Sigma_{(C_2,C_3)}$ becomes
$$
\omega_1|_{\Sigma_{(C_2,C_3)}}=\left(\dfrac{{ \sqrt{2 C_3} e^{-x/2}} \left(C_2 \mathbf{Y}_1(\sqrt{2 C_3} e^{-x/2})+\mathbf{J}_1(\sqrt{2 C_3} e^{-x/2})\right)}{C_2 \mathbf{Y}_0(\sqrt{2 C_3} e^{-x/2})+\mathbf{J}_0(\sqrt{2 C_3} e^{-x/2})}-u\right) dx+du.
$$
A function $I_1=I_1(x,u;C_2,C_3)$ such that $dI_1\wedge \omega_1|_{\Sigma_{(C_2,C_3)}}=0$  is given by
$$
I_1=\sqrt{2 C_3} \psi_{(C_2;C_3)}(x)  +e^{-x}u.
$$
where  
\begin{equation}\label{psi}
    \psi_{(C_2;C_3)}'(x)= \dfrac{C_2 \mathbf{Y}_1(\sqrt{2 C_3} e^{-x/2})+\mathbf{J}_1(\sqrt{2 C_3} e^{-x/2})}{C_2 \mathbf{Y}_0(\sqrt{2 C_3} e^{-x/2})+\mathbf{J}_0(\sqrt{2 C_3} e^{-x/2})}e^{-\frac{3}{2}x}.
\end{equation}


Finally, the solution of \eqref{orden3} is obtained by setting $I_1(x,u;C_2,C_3)=C_1,$ for $C_1\in \mathbb{R},$ which gives
$$
u(x)=-\sqrt{2C_3} e^x \psi_{(C_2;C_3)}(x)  +C_1 e^x,
$$ where the function $\psi_{(C_2;C_3)}$ satisfies \eqref{psi}.

\item {\bf Case II: $C_3<0.$}

In this case a function $I_2=I_2(x,u,u_1;C_3)$ such that $dI_2\wedge \omega_2|_{\Sigma_{(C_3)}}=0$ is given by:
$$ 
I_{2} = \dfrac{e^{x/2} (u_1 - u) \mathbf{I}_0\left(\sqrt{-2 C_3} e^{-x/2}\right) - \sqrt{-2 C_3} \mathbf{I}_1\left(\sqrt{-2 C_3} e^{-x/2}\right)}{e^{x/2} (u - u_1) \mathbf{K}_0\left(\sqrt{-2 C_3} e^{-x/2}\right) - \sqrt{-2 C_3} \mathbf{K}_1\left(\sqrt{-2 C_3} e^{-x/2}\right)} 
$$
where  $\mathbf{I}_{\alpha},\mathbf{K}_{\alpha}$ are the modified Bessel functions  of first and second kind, respectively \cite{handbookfunctions}. 

Proceeding as in the previous case, we obtain the following solution to  equation \eqref{orden3}:
$$
u(x)= \sqrt{-2C_3} e^x \varphi_{(C2;C_3)}(x) + C_1 e^x,
$$ where $$ \varphi_{(C2;C_3)}'(x) =\dfrac{C_2 \mathbf{K}_1(-\sqrt{-2 C_3} e^{-x/2}) - \mathbf{I}_1(\sqrt{-2 C_3} e^{-x/2})}{C_2 \mathbf{K}_0(-\sqrt{-2 C_3} e^{-x/2}) + \mathbf{I}_0(\sqrt{-2 C_3} e^{-x/2})} e^{-\frac{3}{2}x}.$$

\item {\bf Case III: $C_3=0.$}

It can be checked that a solution for the Pfaffian equation defined by  the restriction of the 1-form $\omega_2$ in \eqref{omegas_ej5.1} to the level set  $I_3=0$ is given by 
$$
I_2=\frac{x}{2}+\frac{1}{u_1-u}.
$$
The restriction of the 1-form $\omega_1$ in \eqref{omegas_ej5.1} to the submanifold $\Sigma_{(C_2,0)}$  implicitly defined by $I_3=0,I_2=C_2,$ $C_2\in \mathbb{R}$ becomes 
$$
\omega_1|_{\Sigma_{(C_2,0)}}=-\left(u+\frac{2}{2C_2 - x}\right)dx + du.
$$
The solution of the Pfaffian equation $\omega_1|_{\Sigma_{(C_2,0)}} \equiv 0$ is defined by the function
$$
I_1=u e^{-x}-2e^{-2 C_2}\mbox{E}_1(x-2C_2),
$$
where $\mbox{E}_1=\mbox{E}_1(z)$ denotes the exponential integral function \cite{handbookfunctions}
$$
\mbox{E}_1(z)=\int_z^{\infty} \dfrac{e^{-t}}{t}dt.
$$

By setting $I_1(x,u,C_2)=C_1,$ for $C_1\in \mathbb{R},$ we finally  obtain the following 2-parameter family of exact solutions for equation \eqref{orden3}:
\begin{equation}
\label{sol_exp_int}
u(x)=2 e^{x-2 C_2} \mbox{E}_1(x-2C_2)+C_1e^x.
\end{equation}

\end{enumerate}

\subsection{A fourth-order ODE with a 1-dimensional algebra of Lie point symmetries}
In this subsection we consider  the fourth-order equation
\begin{equation}\label{orden4}
    u_4=u_1 u_3 + (x^2+1)u_2+u_2^2-\frac{1}{2}(x^2+1)u_1^2,
\end{equation}
which has only  the Lie point symmetry $\mathbf{v}=\partial_u$. It can be checked that the Lie reduction method leads to a third-order equation from which it seems difficult to recover the general solution of the initial equation (\ref{orden4}).

By proceeding as in the previous examples,  a \cinf-structure $\langle X_1, X_2,X_3,X_4 \rangle$ for the distribution generated by the vector field
$$
Z=\partial_x+u_1\partial_u+u_2\partial_{u_1}+u_3 \partial_{u_2}+\left(u_1 u_3 + (x^2+1)u_2+u_2^2-\frac{1}{2}(x^2+1)u_1^2\right)\partial_{u_3}
$$
can be explicitly determined by the following vector fields:
\begin{equation}\label{cinf_ej_5_2}
    \begin{array}{l}
X_1=\partial_u,\\
X_2=\partial_{u_1}+u_1\partial_{u_2}+(u_1^2+u_2)\partial_{u_3},\\
X_3=\partial_{u_2}+(x+u_1)\partial_{u_3},\\
X_4=\partial_{u_3}.
\end{array}
\end{equation}

Since $X_1=\partial_u=\mathbf{v}^{(3)},$ where $\mathbf{v}^{(3)}$ denotes the third-order prolongation of the Lie point symmetry $\mathbf{v}$ \cite{olver86}, it is clear that  $X_1$ is a \cinf-symmetry of $\mathcal{S}(\{Z\})$ in the sense of Definition \ref{Csymmdistribution}. The vector field $X_2$ is a \cinf-symmetry of $\mathcal{S}(\{Z,X_1\})$ because 
$$
\begin{array}{l}
[X_2,Z]=X_1+u_1 X_2,\\

[X_2,X_1]=0.\\
\end{array}
$$
The vector field $X_3$ is a \cinf-symmetry of $\mathcal{S}(\{Z,X_1,X_2\})$, since
$$
\begin{array}{l}
[X_3,Z]=X_2+x X_3, \\

[X_3,X_1]=0,\\

[X_3,X_2]=0.\\
\end{array}
$$
Finally, $X_4$ is a \cinf-symmetry of $\mathcal{S}(\{Z,X_1,X_2,X_3\})$ because  $\{Z,X_1,X_2,X_3\}$ are pointwise linearly independent. In this example, $X_2,X_3$ and $X_4$ do not correspond to symmetries of equation \eqref{orden4}.

We use the volume form $\boldsymbol{\Omega}=dx\wedge du\wedge du_1\wedge du_2\wedge du_3$ to calculate 1-forms given by \eqref{formas}:
$$
\begin{array}{l}
\omega_1=u_1 dx-du,\\
\omega_2= -u_2 dx+du_1,\\
\omega_3=-(u_1 u_2-u_3)dx+u_1 du_1-du_2,\\
\omega_4=(-u_2(x^2+xu_1+1)+\frac{1}{2}(x^2 +1)u_1^2+x u_3) dx+(x u_1-u_2)du_1-(x+u_1)du_2+du_3.
\end{array}
$$
\begin{enumerate}
    \item We begin by solving the Pfaffian equation $\omega_4\equiv 0$. It can be checked that a smooth function $I_4=I_4(x,u,u_1,u_2,u_3)$ such that $dI_4 \wedge \omega_4=0$ is given by:
$$
I_4=\left(\frac{1}{2}x u_1^2-u_1 u_2 -x u_2+u_3\right)e^{\frac{1}{2}x^2}.
$$
\item The restriction of $\omega_3$ to the submanifold $\Sigma_{(C_4)}$ implicitly defined by $I_4=C_4,$ $C_4\in \mathbb{R},$ becomes $$
\omega_3 |_{\Sigma_{(C_4 )}}=\left(-\frac{1}{2}x u_1^2+x u_2+C_4 e^{-\frac{1}{2}x^2}\right)dx+u_1 du_1-du_2.
$$   A smooth function $I_3=I_3(x,u,u_1,u_2;C_4)$ such that $dI_3 \wedge \omega_3 |_{\Sigma_{(C_4 )}}=0$ can be expressed in the form:
$$
I_3=-\frac{1}{2} C_4 \sqrt{\pi} \mbox{Erf}(x)+\left( u_2-\frac{1}{2}u_1^2\right)e^{-\frac{1}{2}x^2},
$$
where $\mbox{Erf}=\mbox{Erf}(z)$ denotes the error function defined by \cite{handbookfunctions} 
\begin{equation}
    \label{errorfunc}
    \mbox{Erf}(z)=\dfrac{2}{\sqrt{\pi}}\int_0^{z}e^{-t^2}dt.
\end{equation}

\item \sloppy The restriction of $\omega_2$ to the submanifold  $\Sigma_{(C_3,C_4)}$ of $\Sigma_{(C_4)}$  implicitly defined by $I_3(x,u,u_1,u_2;C_4)=C_3,$ where $C_3\in \mathbb{R},$ becomes $$
\omega_{2}|_{\Sigma_{(C_3,C_4)}}=-\left( \frac{1}{2}C_4 \sqrt{\pi} \mbox{Erf}(x) e^{\frac{1}{2}x^2} +\frac{1}{2}u_1^2+C_3 e^{\frac{1}{2}x^2}\right)dx+du_1.
$$  It can be checked that a function $I_2=I_2(x,u,u_1;C_3,C_4)$ such that $dI_2 \wedge \omega_{2}|_{\Sigma_{(C_3,C_4)}} =0 $ is given by 
\begin{equation}
    \label{I2ODE4}
    I_2 = - \frac{u_1 \psi_2(x;C_3,C_4)+2 \psi_2'(x;C_3,C_4)}{u_1 \psi_1(x;C_3,C_4)+2\psi_1'(x;C_3,C_4)},
\end{equation}
where $\psi_1=\psi_1(x;C_3,C_4)$ and $\psi_1=\psi_1(x;C_3,C_4)$  constitute a fundamental set of solutions to the following two-parameter family of Schr\"{o}dinger-type equations:
\begin{equation}
    \label{lin1}
    \psi''(x) = -\frac{1}{2} e^{\frac{1}{2} x^2} \left(\sqrt{\pi} \frac{C_4}{2} \mbox{Erf}(x) +C_3 \right) \psi(x). 
\end{equation}
\item Finally, the restriction of $\omega_1$ to the submanifold $\Sigma_{(C_2,C_3,C_4)}$ of  $\Sigma_{(C_3,C_4)}$ defined by $I_2(x,u,u_1;C_3,C_4)=C_2$, with $C_2 \in \mathbb{R}$, becomes
$$
\omega_1|_{\Sigma_{(C_2,C_3,C_4)}}=\frac{2 \left( C_2 \psi_1'(x;C_3,C_4) + \psi_2'(x;C_3,C_4) \right)}{C_2 \psi_1(x;C_3,C_4)+\psi_2(x;C_3,C_4)}dx+du.
$$ A function $I_1=I_1(x,u;C_2,C_3,C_4)$ such that $dI_1 \wedge \omega_1|_{\Sigma_{(C_2,C_3,C_4)}} =0 $ can be calculated by a simple quadrature and becomes 
$$I_1=u+2 \ln \left( C_2 \psi_1(x;C_3,C_4)+\psi_2(x;C_3,C_4) \right).$$
\end{enumerate}

As a result of the previous procedure of integration, by using the \cinf-structure defined by \eqref{cinf_ej_5_2}, the initial fourth-order equation \eqref{orden4} has been completely integrated. Its general solution can be expressed in terms of  a fundamental set of solutions $\psi_1(x;C_3,C_4)$ and $\psi_2(x;C_3,C_4)$ of  (\ref{lin1}) as follows: 
\begin{equation}
\label{sol_gen_3}
u(x) = -2 \ln \left( C_2 \psi_1(x;C_3,C_4)+\psi_2(x;C_3,C_4) \right) + C_1,
\end{equation}
where $C_i \in \mathbb{R}$ for $i=1,2,3,4.$

\subsubsection{Some particular solutions in terms of elementary functions}

For some particular values of the arbitrary constants in \eqref{sol_gen_3}, the general solution to equation \eqref{orden4} can be expressed in terms of elementary functions. This is the case, for instance, when $C_3=C_4=0.$ For these particular values, the Schr\"{o}dinger-type equation (\ref{lin1}) turns out to be simply $\psi''=0$ and therefore a corresponding fundamental set of solutions is given by $\psi_1(x)=1$ and $\psi_2(x)=x.$ 

In this case, the expression  \eqref{sol_gen_3}  provides the following two-parameter familiy of exact solutions for equation \eqref{orden3}:
$$
u(x)=-2 \ln(x+C_2)+C_1, \quad C_1,C_2\in \mathbb{R}.
$$

\subsection{A fourth-order ODE without Lie point symmetries}

This example illustrates the success of the \cinf-structure-based method in solving ODEs for which the classical Lie method cannot be applied due to the absence of Lie point symmetries in the equation.  This is the case of  the fourth-order ODE
\begin{equation}\label{orden4bis}
u_4=-uu_1+uu_3+3u_1 u_2+x+u_2,
\end{equation}
whose associated vector field is
$$
Z=\partial_x+u_1\partial_u+u_2\partial_{u_1}+u_3 \partial_{u_2}+\left(-uu_1+uu_3+3u_1 u_2+x+u_2\right)\partial_{u_3}.
$$

It can be checked that the determining equations for a Lie point symmetry of equation (\ref{orden4bis}), in the form $v = \xi(x,u) \partial_x+\eta(x,u) \partial_u$, yield the trivial solution $\xi=\eta=0$. Therefore equation (\ref{orden4bis})
does not  admit Lie point symmetries. 

Finding a \cinf structure for $\mathcal{S}(\{Z\})$ can be simplified by assuming that some of the infinitesimals of the corresponding elements are constant or linear with respect to $u_1$ and $u_2$. This is similar to the approach used in the previous examples.  In this way we find the   ordered set $\langle X_1, X_2,X_3, X_4 \rangle$ given by the  following vector fields:  
$$
\begin{array}{l}
X_1=\partial_u+u\partial_{u_1}+(u^2+u_1)\partial_{u_2}+(u^3+3uu_1+u_2)\partial_{u_3},\\
X_2=\partial_{u_1}+u\partial_{u_2}+(u^2+2u_1)\partial_{u_3},\\
X_3=\partial_{u_2}+(u-1)\partial_{u_3},\\
X_4=\partial_{u_3}.
\end{array}
$$
It can be checked that the vector field $X_1$ is a \cinf-symmetry of $\mathcal{S}(\{Z\})$, since $[X_1,Z]= u X_1$. On the other hand, 
the vector field $X_2$ is a symmetry of $\mathcal{S}(\{Z,X_1\})$, because
$$
\begin{array}{l}
[X_2,Z]=X_1,\\

[X_2,X_1]=0.\\
\end{array}
$$
Finally, $X_3$ is a \cinf-symmetry of $\mathcal{S}(\{Z,X_1,X_2\})$, since the following commutation relations are satisfied:
$$
\begin{array}{l}
[X_3,Z]=X_2- X_3, \\

[X_3,X_1]=0,\\

[X_3,X_2]=0.\\
\end{array}
$$
Thus, in accordance with Definition \ref{Qsolvable}, and considering the pointwise linear independence of $X_1, X_2, X_3, X_4$, the ordered set $\langle X_1, X_2, X_3, X_4 \rangle$ forms a \cinf-structure for $\mathcal{S}({Z})$. 

In what follows, we employ the integration method outlined in Subsection \ref{procedimiento} to achieve our objective of solving  equation (\ref{orden4bis}).
The corresponding 1-forms provided  in \eqref{formas} yield the following expression  when using  $\boldsymbol{\Omega}=dx \wedge du\wedge du_1\wedge du_2\wedge du_3:$ 
\begin{equation}
\label{formas_ej_4bis}
\begin{array}{cl}
\omega_1=&u_1 dx-du,\\
\omega_2=&(uu_1-u_2)dx-udu+du_1,\\
\omega_3=&-(uu_2+u_1^2-u_3)dx+u_1du+udu_1-du_2,\\
\omega_4=&-(-uu_1-uu_2-u_1^2+x+u_2+u_3)dx-(u_1+u_2)du-(u+2u_1)du_1\\ & +(1-u)du_2 +du_3.
\end{array}
\end{equation}

The results obtained after applying the integration procedure, as described in Subsection \ref{procedimiento}, are presented below. 

\begin{enumerate}
\item \sloppy The Pfaffian equation $\omega_4\equiv 0$ is completely integrable and  a function $I_4=I_4(x,u,u_1,u_2,u_3)$ such that $dI_4 \wedge \omega_4 =0$ can be chosen as
$$
I_4=(-u_1^2-u_1u-(u-1)u_2+x+u_3+1)e^{-x}.
$$

\item The restriction of the 1-form $\omega_3$ given in (\ref{formas_ej_4bis}) to $\Sigma_{(C_4)}$ provides  
$$
\omega_3 |_{\Sigma_{(C_4 )}}= (C_4e^{x}+u_1u-x-u_2-1)dx+u_1du+udu_1-du_2.
$$

A function $I_3=I_3(x,u,u_1,u_2;C_4)$ such that $dI_3 \wedge \omega_3 |_{\Sigma_{(C_4 )}}=0$ is
$$
I_3=-\frac{1}{2}C_4e^{2x}-e^{x}(uu_1-x-u_2).
$$
\item We now restrict the 1-form $\omega_2$ in (\ref{formas_ej_4bis}) to $\Sigma_{(C_3,C_4)},$ resulting in
$$
\omega_{2}|_{\Sigma_{(C_3,C_4)}}=-\left(\frac{1}{2}C_4e^{x}-x+C_3e^{-x}\right)dx-udu+du_1.
$$
A function $I_2=I_2(x,u,u_1;C_3,C_4)$ such that $dI_2 \wedge \omega_3 |_{\Sigma_{(C_3,C_4)}}=0$
can be calculated by simple quadrature:  
$$
I_2=-\frac{1}{2}C_4e^{x}+\frac{1}{2}x^2-\frac{1}{2}u^2+u_1+C_3e^{-x}.
$$

\item Finally, the restriction of the 1-form $\omega_1$ given in (\ref{formas_ej_4bis}) to  $\Sigma_{(C_2,C_3,C_4)}$ turns out to be:
\begin{equation}\label{omega1whit}
\omega_{1}|_{\Sigma_{(C_2,C_3,C_4)}}=\left(\frac{1}{2}u^2 +\frac{1}{2}C_4e^{x}-\frac{1}{2}x^2-C_3e^{-x}+C_2\right)dx-du.    
\end{equation}

The integration of the Pfaffian equation  $\omega_{1}|_{\Sigma_{(C_2,C_3,C_4)}} \equiv 0$ is equivalent to solve the following first-order ODE:
\begin{equation}
    \label{ric}
    u_1 = \frac{1}{2}u^2 +\frac{1}{2}C_4e^{x}-\frac{1}{2}x^2-C_3e^{-x}+C_2,
\end{equation}
which is of Riccati-type. Equation (\ref{ric}) can be mapped into the following Schr\"{o}dinger-type equation by means of the standard transformation $u=-2\psi'(x)/\psi(x)$:
\begin{equation}
\label{linear}
\psi''(x) = \left( \frac{1}{4}x^2 - \frac{1}{4} C_4 e^x + \frac{1}{2} C_3 e^{-x} - \frac{1}{2} C_2 \right) \psi(x).
\end{equation}

Therefore, if $\psi_1=\psi_1(x;C_2,C_3,C_4)$ and $\psi_2=\psi_2(x;C_2,C_3,C_4)$ form a fundamental set of solutions to equation (\ref{linear}), a first integral $I_1=I_1(x,u;C_2,C_3,C_4)$ for  the Riccati equation \eqref{ric} is given by \cite[Proposition 4.1]{adrianJPA2017} 

\begin{equation}
I_1=\dfrac{u\psi_1(x;C_2,C_3,C_4)+2\psi_1'(x;C_2,C_3,C_4)}{u\psi_2(x;C_2,C_3,C_4)+2\psi_2'(x;C_2,C_3,C_4)}\end{equation}

\end{enumerate}
By setting $I_1(x,u;C_2,C_3,C_4)=C_1,$ where $C_1\in \mathbb{R},$ we obtain the general solution for equation (\ref{orden4bis}), expressed in terms of a fundamental set of solutions to equation (\ref{linear}): 
\begin{equation}
    \label{sol_gen}
    u(x)= \frac{-2 \left( C_1 \psi_1'(x;C_2,C_3,C_4) + \psi_2'(x;C_2,C_3,C_4) \right)}{C_1 \psi_1(x;C_2,C_3,C_4) + \psi_2(x;C_2,C_3,C_4)},
\end{equation}
where $C_i \in \mathbb{R}$, for $i=1,2,3,4.$
In consequence, the \cinf-structure approach successfully solves the fourth-order equation (\ref{orden4bis}), despite the absence of Lie point symmetries.

\subsubsection{Some families of exact solutions in terms of special functions}

For particular values of the constants $C_2,C_3$ and $C_4$ appearing in \eqref{linear}, a fundamental set of solutions to the Schr\"{o}dinger-type equation \eqref{linear} can be expressed in terms of well-known special functions.  For instance, when $C_3=C_4=0,$ equation \eqref{linear} becomes
$$ 
\psi''(x) = \left( \frac{1}{4}x^2  - \frac{1}{2} C_2 \right) \psi(x),
$$ 
which admits the following linearly independent solutions:
\begin{equation}\psi_1(x;C_2)=\dfrac{\mathbf{W}_{\frac{1}{4}C_2,\frac{1}{4}}\left( \frac{x^2}{2} \right)}{\sqrt{x}},\qquad 
       \psi_2(x;C_2)=\dfrac{\mathbf{M}_{\frac{1}{4}C_2,\frac{1}{4}}\left( \frac{x^2}{2} \right)}{\sqrt{x}},
\end{equation} 
where  $\mathbf{M}_{\mu,\nu}=\mathbf{M}_{\mu,\nu}(z)$ and $\mathbf{W}_{\mu,\nu}=\mathbf{W}_{\mu,\nu}(z)$ denotes the corresponding Whittaker functions \cite{handbookfunctions}, i.e., two linearly independent solutions to the equation  
$$\phi''(z)+\left(\dfrac{-1}{4}+\dfrac{\mu}{z}+\dfrac{\dfrac{1}{4}-\nu^2}{z^2}\right)\phi(z)=0.$$

Therefore, a two-parameter family of solutions that corresponds to \eqref{sol_gen} when $C_3=C_4=0,$ is given by 
\begin{equation}\label{sol_par1}
u(x)=\frac{-x^2 +C_2+1}{x}+\frac{4C_1\mathbf{W}_{\frac{1}{4}C_2+1,\frac{1}{4}}\left( \frac{1}{2}x^2 \right)
		-\left(C_2+3\right)\mathbf{M}_{\frac{1}{4}C_2+1,\frac{1}{4}}\left( \frac{1}{2}x^2 \right)
	}{x\left(C_1\mathbf{W}_{\frac{1}{4}C_2,\frac{1}{4}}\left( \frac{1}{2}x^2 \right)+\mathbf{M}_{\frac{1}{4}C_2,\frac{1}{4}}\left( \frac{1}{2}x^2 \right)\right)}.
\end{equation}

Since Whittaker functions can be defined in terms of hypergeometric or Kummer functions, the family of solutions \eqref{sol_par1} could have alternatively been expressed using other special functions.  Furthermore, by selecting different values for $C_2$ in \eqref{sol_par1}, we can generate 1-parameter families of solutions that involve various types of special functions, such as the following examples: 

\begin{itemize}
   \item For $C_2=0,$ \eqref{sol_par1} provides the next 1-parameter family of exact solutions 
\begin{equation}\label{solpar_bessel}
u(x) = -x\frac{C_1 \mathbf{I}_{-\frac{3}{4}}\left(\frac{1}{4}x^2\right)-\mathbf{K}_{\frac{3}{4}}\left(\frac{1}{4}x^2\right)}{C_1\mathbf{I}_{\frac{1}{4}}\left(\frac{1}{4}x^2\right)+\mathbf{K}_{\frac{1}{4}} \left(\frac{1}{4}x^2\right)},
\end{equation} where $\mathbf{I}_{\nu}$ and $\mathbf{K}_{\nu}$ denote the  modified Bessel functions of the first and second kinds, respectively. 
    
    \item When $C_2=-1,$ we obtain the following 1-parameter family of exact solutions for equation \eqref{orden4}:
\begin{equation}\label{solpar_error}
u(x) = -x+\frac{4e^{-\frac{1}{2}x^2}}{2C_1-\sqrt{2\pi}\mbox{Erf}\left(\frac{1}{2}\sqrt{2}x\right)},
\end{equation}
where $\mbox{Erf}$ denotes the error function \eqref{errorfunc}.

\end{itemize}

\section{Concluding remark}

In this work it has been demonstrated the effectiveness of  the \cinf-structure procedure as a novel tool to deal with integrability problems in differential equations. By applying the integration method based on  \cinf-structures, several models have been fully integrated, including a Lotka-Volterra model and equations for which the Lie method encounters certain obstacles when trying to  obtain the general solution.

Consequently, \cinf-structures offer significant contributions to solving problems that cannot be solved by classical methods, expanding our understanding and analytical capabilities in tackling intricate mathematical problems.

\section{Acknowledgments}
The authors thank the finantial support from Junta de Andalucía to the research group FQM–377 and from Universidad de Cádiz through "Plan Propio de Estímulo y Apoyo a la Investigación y Transferencia 2022/2023". A. Ruiz and C. Muriel thank the partial financial support by the grant "Operator Theory: an interdisciplinary approach", reference ProyExcel$\_$00780, a project financed in the 2021 call for Grants for Excellence Projects, under a competitive bidding regime, aimed at entities qualified as Agents of the Andalusian Knowledge System, in the scope of the Andalusian Research, Development and Innovation Plan (PAIDI 2020). Counseling of University, Research and Innovation of the Junta de Andaluc\'ia.

\bibliographystyle{unsrt}  
\bibliography{references.bib}
\end{document}